\documentclass[submitting]{nst}
\usepackage[T2A,T1]{fontenc}
\usepackage[russian,english]{babel}
\usepackage{cuted}
\usepackage{flushend}
\usepackage{lipsum}
\usepackage{mathtools}
\usepackage{cuted}
\usepackage{url}
\newcolumntype{C}[1]{>{\centering\arraybackslash}p{#1}}

\usepackage{listings}
\begin{document}

\title{Lambda polarization at Electron-ion collider in China} \thanks{Supported by the National Natural Science Foundation of China (No.12275159, No.12075140, No.12175117), 100 Talents Program of CAS, and Shandong Provincial Natural Science Foundation (No. ZFJH202303)}

\author{Zhaohuizi Ji}
\affiliation{Institute of Frontier and Interdisciplinary Science, Shandong University, Qingdao 266237, China}
\author{Xiaoyan Zhao}
\affiliation{Institute of Frontier and Interdisciplinary Science, Shandong University, Qingdao 266237, China}
\author{Aiqiang Guo}
\affiliation{Institute of Modern Physics of CAS, Lanzhou 730000, China}
\author{Qinghua Xu}
\affiliation{Institute of Frontier and Interdisciplinary Science, Shandong University, Qingdao 266237, China}
\author{Jinlong Zhang}
\affiliation{Institute of Frontier and Interdisciplinary Science, Shandong University, Qingdao 266237, China}
\affiliation{Southern Center for Nuclear-Science Theory (SCNT), Institute of Modern Physics, Chinese Academy of Sciences, Huizhou, China}

\begin{abstract}
Lambda polarization can be measured through its self-analyzing weak decay, making it an ideal candidate for studying spin effects in high energy scatterings. In lepton-nucleon deeply inelastic scatterings (DIS), Lambda polarization measurements can probe the polarized parton distribution functions (PDFs) and the polarized fragmentation functions (FFs). 
One of the most promising facilities for high-energy nuclear physics research is the proposed Electron-ion collider in China (EicC). As a next-generation facility, EicC is set to propel our understandings of nuclear physics to new heights. In this article, we study the Lambda production in electron-proton collision at EicC energy, in particular Lambda's reconstruction based on the performance of the designed EicC detector. In addition, taking spontaneous transverse polarization as an example, we provide a theoretical prediction with statistical projection based on one month of EicC data taking, offering valuable insights into future research prospects.
\end{abstract}

\maketitle

\nolinenumbers

\section{Introduction}
Spin, as a fundamental property of particles, plays a crucial role in the advancement of modern physics. 
A growing number of experimental findings, such as the spontaneous transverse polarization of $\Lambda$ and the proton spin crisis, have made it evident that there is much more to be understood about spin behaviors in high-energy reactions. The Lambda hyperon ($\Lambda/\overline{\Lambda}$) emerges as an exceptionally powerful tool in spin physics, primarily due to its parity-violating weak decay, which results in a non-uniform angular distribution of its products with respect to Lambda's spin direction ~\cite{leeyang1957}. In high energy reactions Lambda can be abundantly produced and efficiently detected via the decay channel $\Lambda \rightarrow p\pi^{-}$ (branching ratio is $63.9\%$). In the $\Lambda$ rest frame, the decay protons are preferentially emitted along the polarization direction of their parent $\Lambda$, with the following angular distribution, 
\begin{equation}
    \frac{dN}{d\cos{\theta^*}} \propto \mathcal{A}(1+\alpha_{\Lambda(\overline{\Lambda})}P_{\Lambda(\overline{\Lambda})}\mathrm{cos}{\theta^*}),
\end{equation}  
where $\mathcal{A}$ is the detector acceptance,  $\alpha_{\Lambda}$ = 0.732$\pm$0.014 is the  weak decay parameter~\cite{pdg2022}, $\theta^{*}$ is the angle between proton momentum direction and $\Lambda(\overline{\Lambda})$ polarization direction in the $\Lambda$ rest frame.

The spontaneous transverse polarization of $\Lambda$ was first observed in 1976 in the unpolarized proton beam scattering on Beryllium target~\cite{bunce1976lambda}, by that time the perturbative Quantum Chromodynamics (QCD) only predicted a negligible polarization~\cite{PhysRevLett.41.1689}. 
This puzzling results triggered a series of  theoretical and phenomenological studies which had been extended far beyond $\Lambda$ polarization itself. 
Experimentally, measurements of $\Lambda$ polarization have since been extensively explored in various high-energy processes, encompassing electron-positron annihilation~\cite{aleph1996,opal1998,belle2019}, lepton-nucleon deeply inelastic scattering (DIS)~\cite{e6652000,hermes2001,nomad2000}, hadron-hadron scattering~\cite{star2009,star2018l, star2018t}, and heavy ion collisions~\cite{star2017nat,star2023nat,wang2023,chen2023,acta2023}, yielding invaluable insights into numerous aspects of physics. These measurements have served diverse purposes, including unraveling the physical origins of spontaneous polarization, understanding nucleon spin structure, comprehending spin effects in fragmentation processes, and exploring extreme conditions of high density and high temperature in heavy ion collisions.

High precision Lambda polarization measurements in the proposed electron-ion collider worldwide, provide unique opportunities to study the spin-dependent fragmentation functions (FFs) and polarized parton distribution functions (PDFs)~\cite{Lu:1995np,Ellis:1995fc,Jaffe:1996wp,Ma:2000uu,Ellis:2002zv,Zhou:2009mx,Chen:2021zrr,Kang:2021kpt}.
The Electron-ion collider in China, EicC, is the proposed next generation high energy nuclear physics facility, which is based on the High Intensity Heavy-ion Accelerator Facility (HIAF) in Huizhou, China~\cite{eiccwp_ch,eiccwp_eng}. 
It is conceptually designed to deliver high luminosity electron-proton, electron-ion collision, with electron, proton, and light ion beams highly polarized. With complementary kinematics coverage to the other electron-ion collider proposals worldwide~\cite{enc2010,lhec2012,useic2022}, the featured physics at EicC includes 3-dimensional proton spin structure, nuclear partonic structure, exotic hadron states, {\it etc}. Lambda polarization measurement at EicC is expected to be sensitive to not only the spin dependent parton distribution functions, but also the spin dependent fragmentation functions. Potential measurements of Lambda transverse polarization and impact studies have been performed for US-based EIC which is designed to collide electron and proton/ion beams at significantly higher energies than EicC~\cite{Kang:2021kpt}.  

In this work, the Lambda production in electron-proton scattering under EicC configuration is studied. Based on the current conceptual EicC detector design, especially the design of  tracking subsystem, the reconstruction performance for $\Lambda/\overline{\Lambda}$ is assessed. In section~\ref{sec:simu}, we will describe the simulation setup including the event generator in use, the detector configuration and the corresponding fast simulation procedure. Performance of $\Lambda/\overline{\Lambda}$ reconstruction will be presented in section~\ref{sec:reco}. In section~\ref{sec:projection}, taking the spontaneous transverse polarization as an example, the potential statistical precision for polarization measurements will be given together with theoretical predictions.  We will give a brief summary and outlook in section~\ref{sec:summary}. 

\section{Simulation Framework}\label{sec:simu}

\begin{figure}[hbt!]
  \centering
  \includegraphics[width=0.33\textwidth]{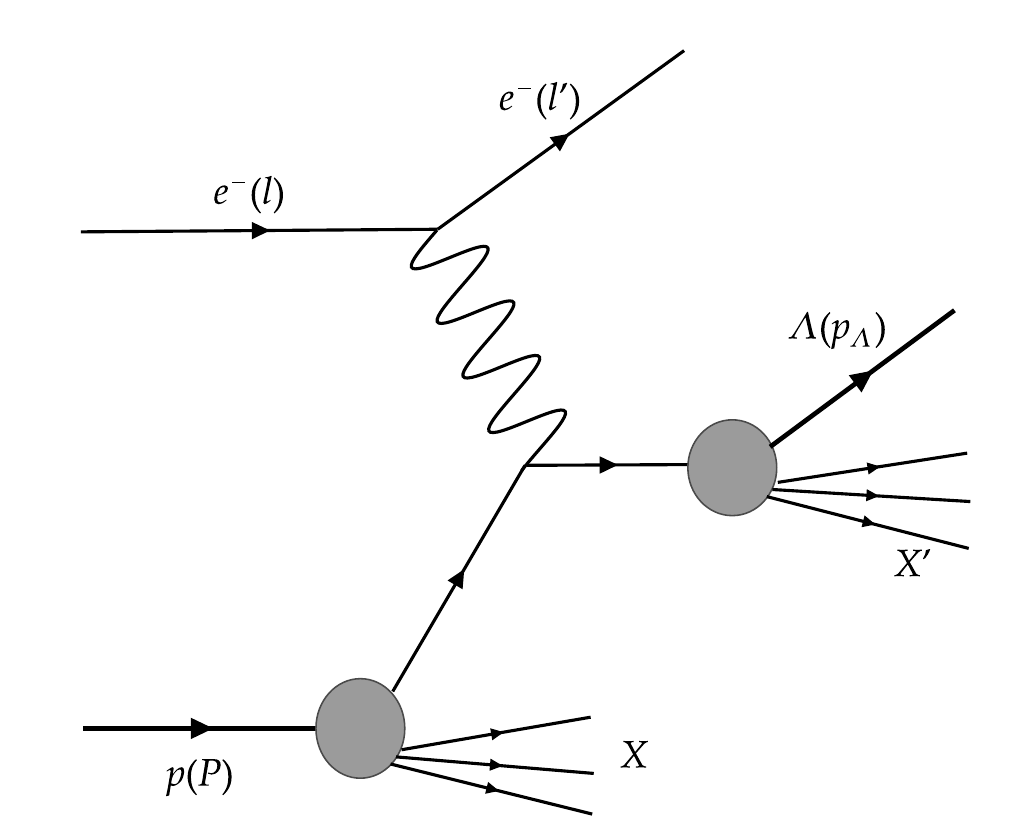}
  \caption{ Leading-order diagram for ${\Lambda}$ production in a semi-inclusive DIS process.}
  \label{fig:SIDIS}
\end{figure}

To simulate the Lambda production in electron-proton scattering, event generator PYTHIAeRHIC~\cite{pythiaerhic}, a modified version of PYTHIA6.4.28~\cite{pythia6}, is used with the Parton Distribution Functions (PDFs) input from LHAPDF~\cite{lhapdf}. The collision energy we choose is the baseline energy outlined in the EicC whitepaper~\cite{eiccwp_eng}, 3.5 GeV electron on 20 GeV proton. The leading-order diagram for $\Lambda$ production in DIS process is shown in Fig.~\ref{fig:SIDIS}. The kinematics of the studied DIS events are constrained in the following ranges: Bjorken-$x$ $10^{-3}<x_{B} < 1$, transferred 4-momentum squared $Q^2 > 1$ GeV$^2$, and hadronic invariant mass squared $W^2 > 4$ GeV$^2$.  10 millions of such DIS events are generated for the following studies.  

\begin{figure*}[bht!]
    \centering
    \includegraphics[width=0.92\textwidth]{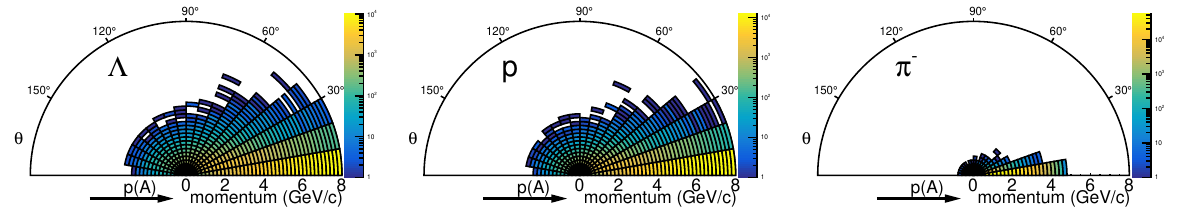}
    \caption {Momentum (radial) and polar angle (polar) distributions for $\Lambda$ and its decay products in the laboratory frame.} 
    \label{fig:kin_map}
\end{figure*}

At generator level, the average number of $\Lambda$ produced per DIS event in the above kinematics ranges is about 0.1. In the laboratory frame, the momentum and polar angle distributions for $\Lambda$ and the decay products are shown in Fig.~\ref{fig:kin_map}. Comparing distributions of daughter proton and pion with of $\Lambda$, it can be found that proton carries most of $\Lambda$'s momentum while pion only shares a small fraction. $\Lambda$ is preferentially produced in proton going direction and with a large amount produced at very forward angle. Same distributions in Fig.~\ref{fig:kin_map} for $\overline{\Lambda}$ are similar with slight difference which will be discussed in later context. 

\begin{figure}[bht!]
    \centering  \includegraphics[width=0.43\textwidth]{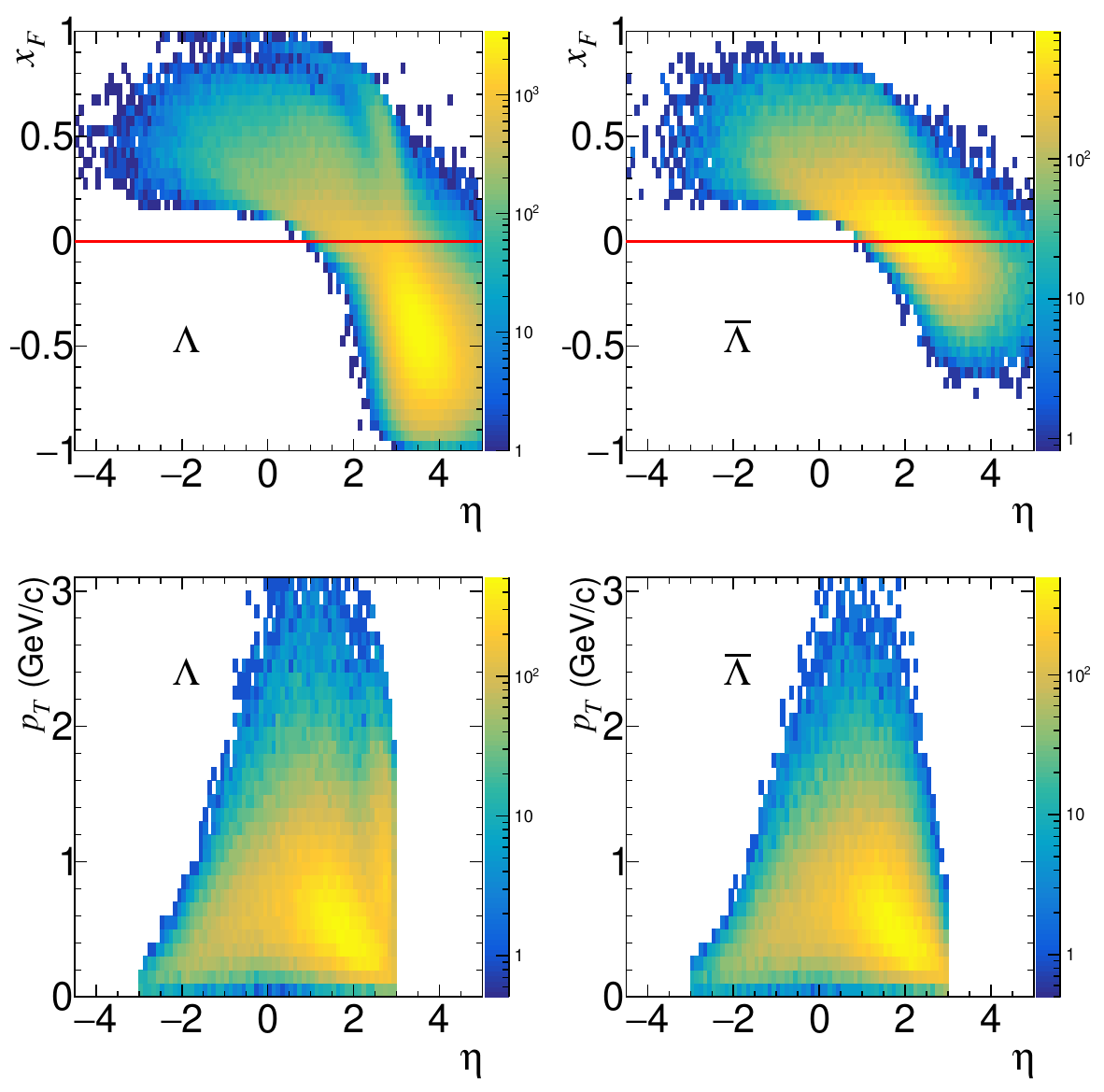}
    \caption{Upper panels: Feynman-$x$ $x_F$ vs. pseudorapidity $\eta$ for $\Lambda$ (left) and $\overline{\Lambda}$ (right) in the laboratory frame. Only $\Lambda/\overline{\Lambda}$ above the red line ($x_F > 0$) are kept. Lower panels: transverse momentum $p_T$ vs. pseudorapdity $\eta$ for $\Lambda$ (left) and $\overline{\Lambda}$ (right) with $x_F > 0$ and $|\eta|<3$ (also $|\eta|<3$ for daughter proton and pion). }
    \label{fig:xf_eta}
\end{figure}

\begin{figure}[bht!]
    \centering
    \includegraphics[width=0.36\textwidth]{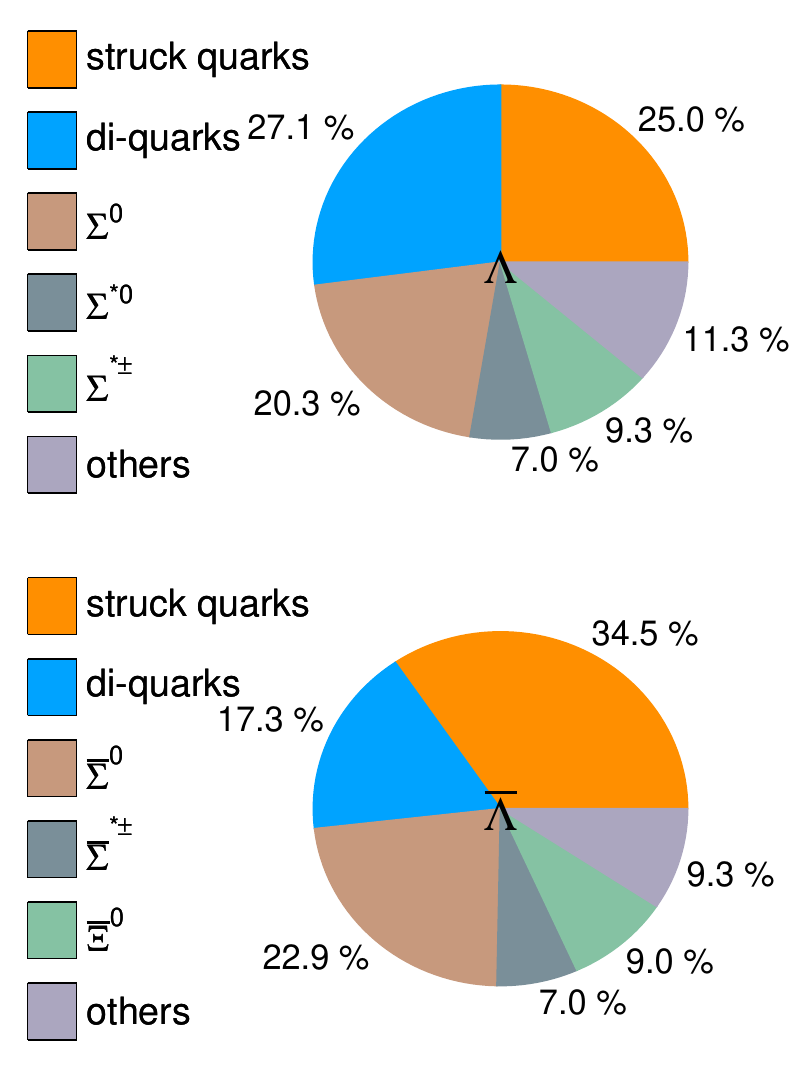}
    \caption{Origins of $\Lambda$ and $\overline{\Lambda}$ with $x_F > 0$, $|\eta|<3$ based on PYTHIA event record.}
    \label{fig:origin}
\end{figure}

In this work, we are mostly interested at the $\Lambda/\overline{\Lambda}$ from the struck quark fragmentation (current fragmentation region). Typically, Feynman-x is an effective variable to separate current fragmentation region and target fragmentation region. 
Feynman-$x$ $x_F$ is defined as $x_F\equiv 2 p_{L}^{\Lambda(\overline{\Lambda})}/W$, where $p_{L}^{\Lambda(\overline{\Lambda})}$ is the $\Lambda(\overline{\Lambda})$ longitudinal momentum in the hadronic center of mass frame, and $W$ the hadronic invariant mass.
Criteria $x_{F} > 0$ is expected to suppress the contributions from target fragmentation region. The correlation between Feynman-$x$ and $\Lambda/\overline{\Lambda}$ pseudorapidity $\eta$ are shown in upper panels in Fig.~\ref{fig:xf_eta}. Here, pseudorapidity is defined as $\eta=-\mathrm{ln}(\mathrm{tan}(\theta/2))$, where $\theta$ is the polar angle. Following the EicC convention, the positive $\eta$ is along the moving direction of proton/ion beam.
One can see that $\Lambda/\overline{\Lambda}$ with $x_F<0$ are mostly produced at very forward region which will be discarded in the following simulation and analysis. Considering the limited coverage of EicC central detector, $|\eta|<3$ is applied for $\Lambda/\overline{\Lambda}$ and their daughter particles. Transverse momentum, $p_T$, vs. $\eta$ for $\Lambda$ and $\overline{\Lambda}$, with $x_F > 0$ and $|\eta| < 3$ are shown in the lower panels of Fig.~\ref{fig:xf_eta}.  By tracing back the full event records in PYTHIA, the origins of such $\Lambda/\overline{\Lambda}$ are shown in Fig.~\ref{fig:origin}. At EicC energy, about half of $\Lambda/\overline{\Lambda}$ are from decay of heavier hyperons. There are also significant contribution from beam remnants (di-quarks). In this study, we don't separate different sources for $\Lambda/\overline{\Lambda}$. 

The preliminary conceptual design of EicC detector has been described in the white papers~\cite{eiccwp_ch,eiccwp_eng}. From inner to outer, it consists of the vertex/tracking detector, the particle identification (PID) system , the calorimeter system, {\it etc}. For $\Lambda$ measurement, the most relevant parts are the tracking and PID systems. The latest design of the EicC tracking detector is thoroughly described in Ref.~\cite{eicc_hf2023}. 
The current design of tracking system uses hybrid models. For middle rapidity ($|\eta|<1.1$), there are 5 layers of silicon and 4 layers of Micro-Pattern Gaseous Detectors (MPGD), radially ranging from 3.3 cm to 77.5 cm. For $|\eta|>1.1$, the tracking system consists of silicon disks followed by large-area Micromegas in the forward (proton/nucleus going) direction and all silicon disks in the backward (electron-going) direction. For the PID system, time of flight detector and Cherenkov detector will be used for particle identification at middle and forward rapidity respectively.

For the tracking system, full GEANT4 simulation has been performed with the latest design, based on which the resolutions for primary vertex position, for distance from tracks to tracks and from tracks to points, and for track momentum, the tracking detector efficiency as function of track $p_T$ and $\eta$  are given in details in Ref.~\cite{eicc_hf2023} (Fig.4-9 therein).  
As well, a fast simulation framework is also developed to simulate the detector responses learned from the GEANT4-base simulation. In this work, we follow the same fast simulation procedure described in Ref.~\cite{eicc_hf2023}. 
The detailed GEANT4 simulation for PID system is not available when this work is performed. To mimic the particle identification imperfection, a simplified “PID smearing” is included in the detector effect fast simulation. In principal the PID efficiency is correlated with the momentum of particles. However, we employ a toy model to study the PID effect with a typical PID efficiency of 95\% as the following. Basically the identified $\pi$, $K$, or $p$ has $95\%$ possibilities to be correct, and $2.5\%$ possibilities to be one of other two particles respectively, as described by the following matrix:

\begin{gather}
 \begin{bmatrix} \pi \\ K  \\ p \end{bmatrix} _{\rm{smeared}}
 =
  \begin{bmatrix}
   0.95 & 0.025 & 0.025 \\
   0.025 & 0.95 & 0.025\\
   0.025 & 0.025 & 0.95
   \end{bmatrix}
   \begin{bmatrix} \pi \\ K  \\ p \end{bmatrix} _{\rm{truth}}
\end{gather}

\noindent Here $95\%$ of the PID purity is specifically chosen, and a few other numbers are also checked for a complete study. 

\section{Lambda reconstruction}\label{sec:reco}
\begin{figure}[hbt!]
  \centering
  \includegraphics[width=0.38\textwidth]{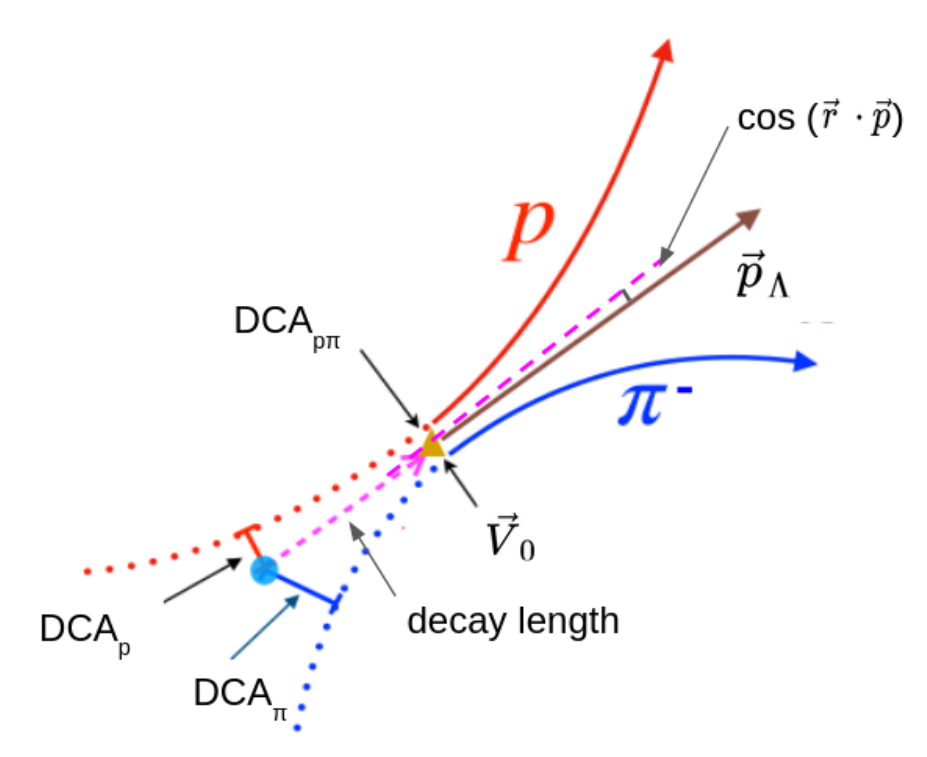}
  \caption{Topology schematic diagram of $\Lambda$ production and its decay process through $\Lambda \rightarrow p\pi^{-}$.}
  \label{fig:topo}
\end{figure}

Similar to the method used in other experiments with tracking detector, $\Lambda/\overline{\Lambda}$ reconstruction in this study is based on the topological structure of the decay channel with a large branching ratio, $\Lambda \rightarrow p\pi^{-}$ and $\overline{\Lambda} \rightarrow \overline{p} \pi^+$. Taking $\Lambda$ as example, Fig~\ref{fig:topo} schematizes the main topological features of its production and decay process in a tracking detector.
Blue dot at the bottom-left represents the $ep$ scattering vertex, named as “primary vertex”. $\Lambda/\overline{\Lambda}$ is emitted from the primary vertex, then moves along the magenta straight dash line and decay at “V0 vertex”. The decay products $p\pi^{-}$($\overline{p}\pi^+$) travel along helical lines with opposite bending directions in the magnetic field.

Reconstruction of $\Lambda$/$\overline{\Lambda}$ starts with pairing proton and pion tracks with opposite charge. To select the  $\Lambda$/$\overline{\Lambda}$ candidates and suppress the random backgrounds, the following selection variables are considered : 

(1) The distance of closest approach (DCA) of proton and pion tracks to the primary vertex. As indicated in Fig.~\ref{fig:topo}, DCA$_p$ and DCA$_{\pi}$ from signals should be significantly higher than those from background, as parent $\Lambda$/$\overline{\Lambda}$ flies certain distances from primary vertex before its decay. 

(2) The distance of closest approach (DCA) between paired proton tracks and pion tracks. For $\Lambda$/$\overline{\Lambda}$ signal, this variable should be consistent with zero within the track space resolution. The decay point (V0 vertex) is given by the middle point of these two tracks at the closest approach, as indicated by the brown triangle in Fig.~\ref{fig:topo}. 

(3) The decay length of $\Lambda/\overline{\Lambda}$ candidates, which is the distance between the primary vertex and V0 vertex. The characteristic decay length of $\Lambda$ hyperon $c\tau$ is 7.89 cm \cite{pdg2022}. 

(4) The angle between $\Lambda/\overline{\Lambda}$ candidate momentum $\Vec{p}$ and its trajectory $\Vec{r}$ from primary vertex. For the $\Lambda/\overline{\Lambda}$ directly produced from the primary vertex, its momentum direction is supposed to be along its trajectory from primary vertex. Correspondingly, cos($\Vec{r}\cdot\Vec{p}$) should be very close to 1. 

\begin{figure*}[t]
  \centering
  \includegraphics[width=0.90\textwidth]{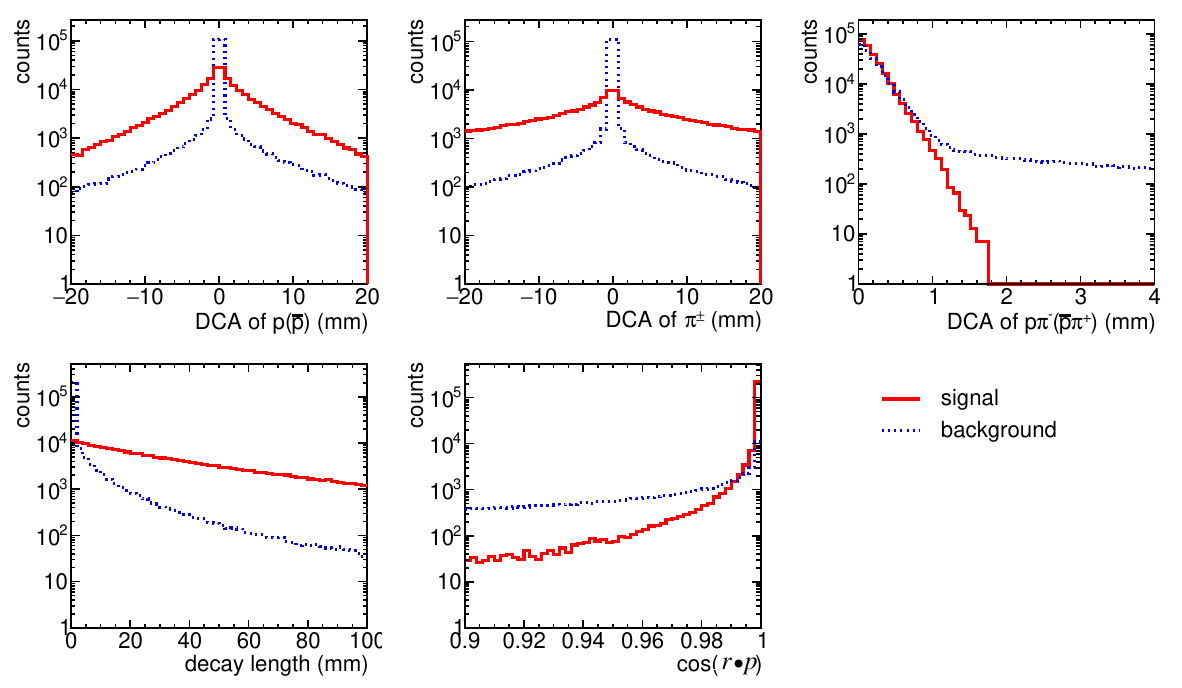}
  \caption{Distributions of topological variables for $\Lambda/\overline{\Lambda}$ signal (red) and background (blue) respectively.}
  \label{fig:cuts}
\end{figure*}

To quantitatively determine the selection criteria, the distributions of proton-pion pairs from pure $\Lambda/\overline{\Lambda}$ sample are compared with the proton-pion pairs from backgrounds, and the  
comparisons are shown in Fig.~\ref{fig:cuts}. 
Based on the comparisons, a set of selection criteria are optimized to balance the background fraction and the $\Lambda/\overline{\Lambda}$ reconstruction efficiency,  in order to keep as many signals as possible while keeping the background fraction at a relatively low level. The numerical cut conditions are listed in Tab.~\ref{tab:cuts}.

\begin{table}[h]
\centering
 \begin{tabular}{ C{0.2\textwidth} C{0.23\textwidth}} 
 \hline
 Variables & Cut condition \\
 \hline
 DCA$_p$  &  $> 0.1$ mm\\
 DCA$_{\pi}$ & $> 0.5$ mm\\
 DCA of $p\pi$ pair & $< 0.8$ mm \\ 
 Decay length & $> 1.5$ mm \\ $\cos{(\Vec{r}\cdot\Vec{p})}$ & $> 0.95$\\  
 \hline
 \end{tabular}
 \caption{The summary of topological criteria for $\Lambda/\overline{\Lambda}$ reconstruction. }
\label{tab:cuts}
\end{table}

\begin{figure}[hbt!]
  \centering
  \includegraphics[width=0.35\textwidth]{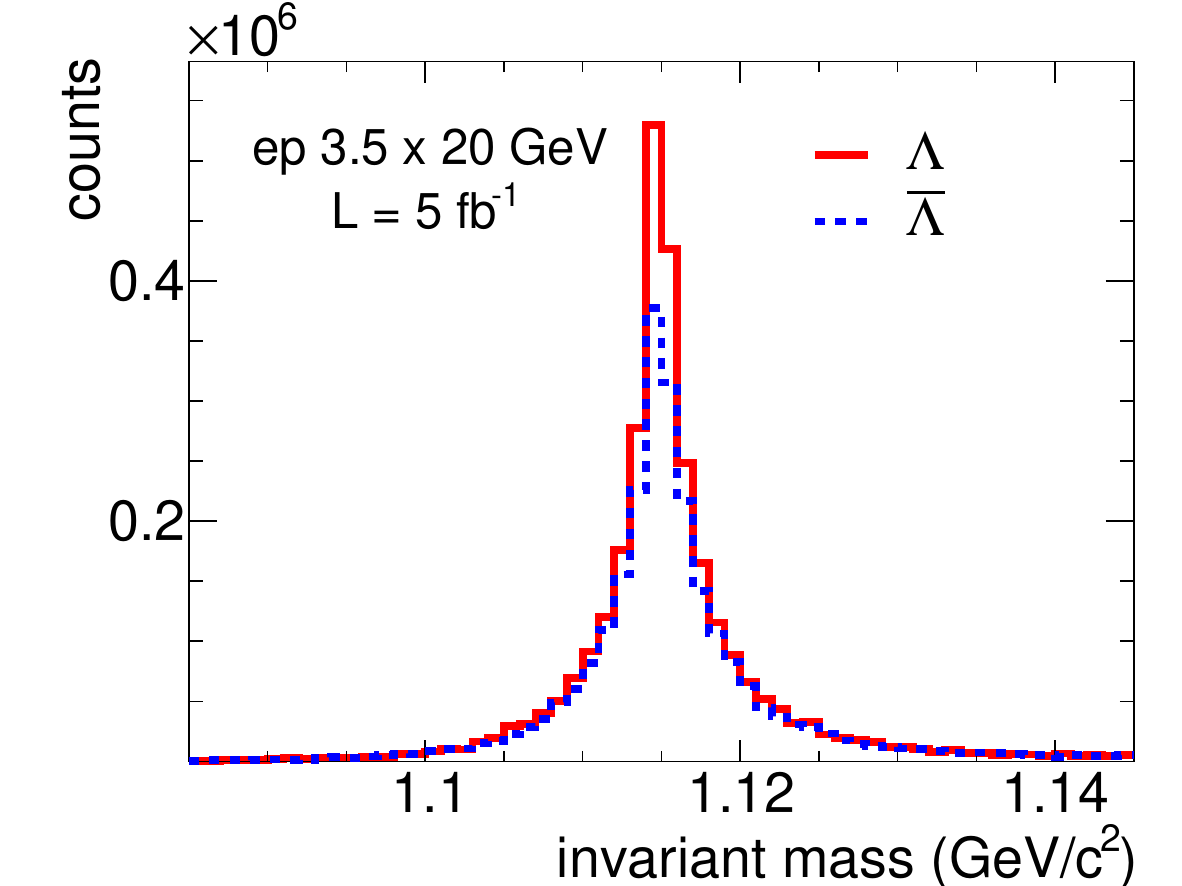}
  \caption{ Invariant mass distributions of $\Lambda$ and $\overline{\Lambda}$ candidates passing all selection criteria.}
  \label{fig:mass}
\end{figure}

By implementing the aforementioned selection criteria, we successfully obtained a clean sample of $\Lambda/\overline{\Lambda}$ candidates. The invariant mass spectrum of $\Lambda/\overline{\Lambda}$ candidates with kinematics cuts of  $x_{F} > 0$, $|\eta|<3$, and $z_{\Lambda} > 0.1$ (fractional momentum of $\Lambda/\overline{\Lambda}$ is defined as $z_{\Lambda}\equiv\frac{P \cdot p_\Lambda}{P \cdot q}$ ) is shown in Fig.~\ref{fig:mass}. Here the shown histograms are scaled to an integral luminosity of 5 fb$^{-1}$ which is corresponding to about 1 month of EicC data taking. 
With all selection criteria applied, more $\Lambda$ over $\overline{\Lambda}$ are reconstructed due to the baryon number enhancement. There is a clean Gaussian signal peak with very limited background. The residual background mainly comes from the random combinations of oppositely charged particles and particle mis-identification. 
Invariant mass distribution of these background is expected to be linear. With typical side-band method, the residual background fraction is estimated to be about $2.6\%$ for $\Lambda$ and $3.0\%$ for $\overline{\Lambda}$. Signal mass window is set to be within $3\sigma$ width of Gaussian fit, which is $(1.106,1.124)$ GeV/c$^2$. 
As for the side bands, they are limited to the regions far from the mass window to avoid fluctuations by signals, but not far enough to escape from the signal peak. The left side band is $(1.083, 1.093)$ GeV/c$^2$ and right side band is $(1.137, 1.147)$ GeV/c$^2$. 
The background under the signal peak is estimated as the sum of the two sides-bands normalized to the signal window.
As mentioned in section~\ref{sec:simu}, the sensitivity of $\Lambda/\overline{\Lambda}$ reconstruction to the PID performance is assessed by varying the PID "purity" number. For $100\%$ PID purity, the residual background fraction is $1.7\%$, while for a $90\%$ case, the residual background fraction increases to $3.4\%$, which is still under good control.

\begin{figure}[hbt!]
     \centering     
     \includegraphics[width=0.35\textwidth]{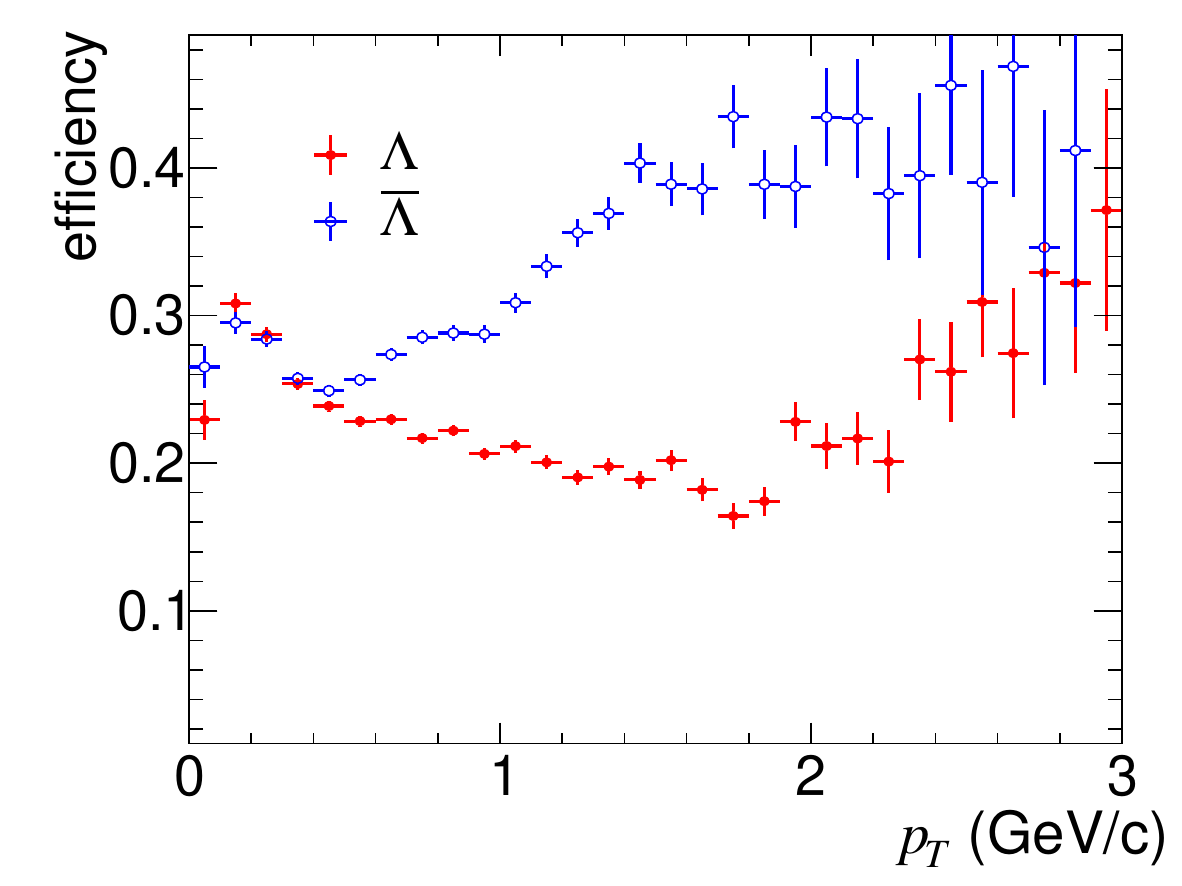}
     \caption{ Reconstruction efficiency of $\Lambda$ and $\overline{\Lambda}$ as a function of $p_{T}$ after all selection criteria applied.}
        \label{fig:eff}
\end{figure}

Figure~\ref{fig:eff} shows the $\Lambda$ and $\overline{\Lambda}$ reconstruction efficiency versus transverse momentum. The reconstruction efficiency involves several effects including topological cuts, detector acceptance, track efficiency which depends on track $p_T$, track $\eta$, etc. For $\Lambda/\overline{\Lambda}$ with large decay length, the number of tracking detector layers the daughter tracks pass through decreases and so does the tracking efficiency. The efficiency at very low $p_T$ is limited by the  detector acceptance due to magnetic field. 
Due to low transverse momenta in the forward region (large $|\eta|$), the efficiency decreases significantly. 
As already shown in Fig.~\ref{fig:xf_eta}, more $\Lambda$ than $\overline{\Lambda}$ are produced at large pseudorapidity where the reconstruction efficiency is low, which lead to significantly higher efficiency of $\overline{\Lambda}$ compared $\Lambda$ to at $p_T>0.5$ GeV/c. When $p_T$ is larger than 2 GeV/c, efficiency for $\Lambda$ reconstruction increases and gets closer to $\overline{\Lambda}$ since $\Lambda$ at middle rapidity start to dominate its production. 

\section{Spontaneous transverse polarization}\label{sec:projection}
In this section, we take $\Lambda/\overline{\Lambda}$ spontaneous transverse polarization as an example to explore the physics potentials of EicC. The theoretical calculation and statistical projection based on our simulation results are described in the following.  

The QCD formalism is used in describing $\Lambda$ spontaneous transverse polarization $P_{\Lambda}$ in the semi-
inclusive DIS process, $e^-(l)+p(P)\xrightarrow{}e^-(l')+\Lambda(p_\Lambda,\mathbf{S}_{\Lambda \perp})+ X$.
The Trento convention~\cite{Bacchetta:2004jz} is followed in the calculation, where the virtual photon moves in the positive $z$ direction and the proton moves in the negative $z$ direction, and the differential cross section can be written as~\cite{Boer:1997nt,Bacchetta:2006tn},
\begin{equation}
   \label{eq:structure function} 
	\begin{aligned}
         &\frac{d\sigma(\mathbf{S}_{\Lambda \perp})}{dx_Bdydz_\Lambda d^2\boldsymbol{p}_{\Lambda 	\perp}} =\frac{4\pi\alpha_{em}^{2}}{yQ^{2}}\frac{y^2}{2(1-\epsilon)}\left(1+\frac{\gamma^2}{2x_B} \right)\left\lbrace  F_{UU} \right.
          \\ 
         &\ \ \ \ \ \ \ \ \ \ \ \ \ \ \ +\left.  \left| \mathbf{S}_{\Lambda \perp}\right| \sin(\phi_{S_\Lambda}-\phi_\Lambda) F^{\sin(\phi_{S_\Lambda}-\phi_\Lambda)}_{UT}+\cdot\cdot\cdot\right\rbrace,  
    \end{aligned}	
\end{equation}
where $\gamma=2x_BM/Q$ and $Q^2=-q^2$, $x_B=\frac{Q^2}{2P \cdot q}$, $y=\frac{P \cdot q}{P \cdot l}$, $z_\Lambda=\frac{P \cdot p_\Lambda}{P \cdot q}$ are Lorentz invariant variables, and $\mathbf{S}_{\Lambda \perp},\ \boldsymbol{p}_{\Lambda\perp}$ are the transverse spin vector and transverse momentum of the $\Lambda$ hyperon, respectively. 
$F_{AB}=F_{AB}(x_B,z_\Lambda,\boldsymbol{p}_{\Lambda\perp},Q)$, where the subscripts indicate the polarization of proton and $\Lambda$, respectively. 
$F_{UU}$ is the spin-averaged structure function, and $F^{\sin(\phi_{S_\Lambda}-\phi_\Lambda)}_{UT}$ is the spin-dependent term that contributes to the spontaneous transverse polarization.
The experimentally measured polarization $P_{\Lambda}$ is related to the structure functions as follows: 
\begin{equation}\label{eq:Polarization}
	\begin{aligned}
		P_{\Lambda}=\frac{F^{\sin(\phi_{S_\Lambda}-\phi_\Lambda)}_{UT}}{F_{UU}}.
	\end{aligned}	
\end{equation}
Within the usual transverse momentum distribution (TMD) factorization, at leading twist the structure functions can be written as,
\begin{equation}
	\begin{aligned}
	   &F_{UU}=\int d^2\boldsymbol{p}_{\perp}d^2\boldsymbol{k}_{\perp}\delta^2(z_\Lambda\boldsymbol{p}_{\perp}+\boldsymbol{k}_{\perp}-\boldsymbol{p}_{\Lambda\perp}) \\
        &\ \ \ \ \ \ \ \ \ \ \ \ \ \times\sum_{q}e_{q}^{2}f_{1q}(x_B,\boldsymbol{p}_{\perp}^2,Q)D_{1q}^\Lambda(z_\Lambda,\boldsymbol{k}_{\perp}^2,Q),\\
		&F^{\sin(\phi_{S_\Lambda}-\phi_\Lambda)}_{UT}=\int d^2\boldsymbol{p}_{\perp}d^2\boldsymbol{k}_{\perp}\delta^2(z_\Lambda\boldsymbol{p}_{\perp}+\boldsymbol{k}_{\perp} -\boldsymbol{p}_{\Lambda\perp}) \\
        &\ \ \ \ \ \times\sum_{q}e_{q}^{2} \frac{\hat{\boldsymbol{p}}_{\Lambda\perp}\cdot\boldsymbol{k}_\perp}{z_\Lambda M_\Lambda} f_{1q}(x_B,\boldsymbol{p}_{\perp}^2,Q)D_{1Tq}^{\perp\Lambda}(z_\Lambda,\boldsymbol{k}_{\perp}^2,Q),
	\end{aligned}	
\end{equation}
where $\boldsymbol{p}_{\perp}$ and $\boldsymbol{k}_{\perp}$ denote the transverse momentum of the quark relative to the initial proton and the transverse momentum of $\Lambda$ relative to its parent quark, respectively.

\begin{figure*}[tbh!]
    \includegraphics[width=\textwidth]{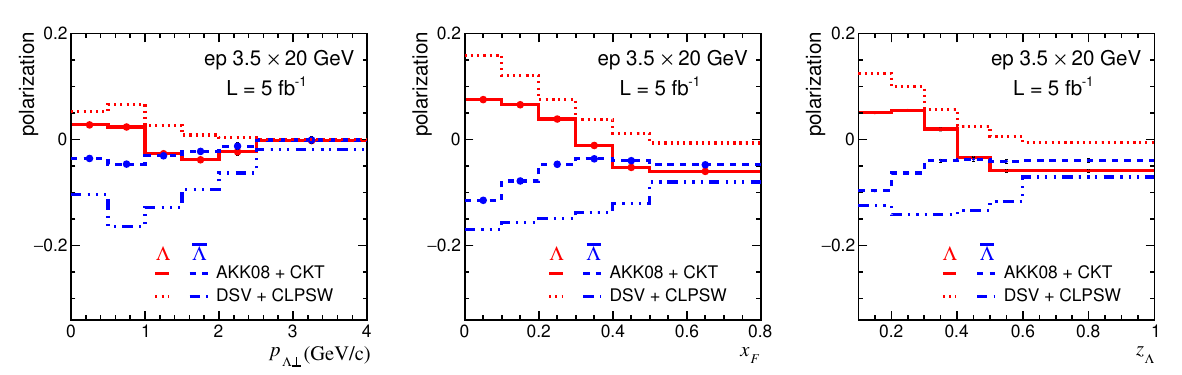}
    \caption{The statistical projection with theoretical predictions for $\Lambda$ and $\overline{\Lambda}$ polarization in $ep$ collisions at EicC. The size of the projected statistical errors are smaller than the marker sizes, thus invisible.}
    \label{fig:polarization}
\end{figure*}

We parameterize the TMDs using the usual Gaussian form, as the product of collinear functions and Gaussian widths:
\begin{equation}\label{eq:Ffunctions}
	\begin{aligned}
		&f_{1q}(x_B,\boldsymbol{p}^2_{\perp};Q)=f_{1q}(x_B,Q)\frac{e^{-\boldsymbol p^2_{\perp}/ \left\langle p_{\perp}^2 \right\rangle }}{\pi\left\langle p_{\perp}^2 \right\rangle }, \\ &D_{1q}^\Lambda(z_\Lambda,\boldsymbol k^2_{\perp};Q)=D_{1q}^\Lambda(z_\Lambda,Q)\frac{e^{-\boldsymbol k^2_{\perp}/\left\langle k_{\perp}^2 \right\rangle }}{{\pi\left\langle k_{\perp}^2 \right\rangle }}, \\ 
		&D_{1Tq}^{\perp\Lambda}(z_\Lambda,\boldsymbol k^2_{\perp};Q)=D_{1Tq}^{\perp\Lambda}(z_\Lambda,Q)\frac{e^{-\boldsymbol k^2_{\perp}/\left\langle M_D^2 \right\rangle }}{{\pi \left\langle M_D^2\right\rangle  } },
	\end{aligned}
\end{equation}
where $\left\langle p_{\perp}^2 \right\rangle =0.61,\left\langle k_{\perp}^2 \right\rangle =0.19$ and $\left\langle M_D^2 \right\rangle =0.118$ are the corresponding Gaussian widths from Ref.~\cite{Anselmino:2013lza,Callos:2020qtu}.
In this analysis, we use the CT18NLO~\cite{Hou:2019qau} parametrization for the collinear PDF while using the DSV~\cite{deFlorian:1997zj} and AKK08~\cite{Albino:2008fy} parametrizations for the collinear unpolarized FF. 
Both parametrizations describe experimental data but significantly differ, with AKK08 including substantial isospin symmetry violations and DSV parametrization conserving it.
Additionally, the universality of the polarizing FF $D_{1Tq}^{\perp\Lambda}$ has been proven~\cite{Metz:2002iz,Meissner:2008yf,Boer:2010ya}. 
Similarly, $D_{1Tq}^{\perp\Lambda}$, as a modulation of $D_{1q}^\Lambda$ by an additional collinear function, is also described by two different parametrizations, i.e. CLPSW~\cite{Chen:2021hdn} considering isospin symmetry and CKT allowing isospin symmetry violations~\cite{Callos:2020qtu}. 
The future EicC experiment provide an ideal place to test the $\Lambda$ isospin symmetry of FFs.
In our study, we employ these different parametrizations to calculate the polarization observables and perform comparisons.

Using these parametrizations for the TMDs in Eq.(\ref{eq:Ffunctions}), the spontaneous transverse polarization of $\Lambda$ in Eq.(\ref{eq:Polarization}) has the analytic form:

\begin{equation}\label{eq:PA1}
	\begin{aligned}
		&P_{\Lambda}(x_B,z_\Lambda,\boldsymbol p_{\Lambda\perp},Q)\\
        &=\frac{\sum_{q}e_{q}^{2}f_{1q}(x_B,Q)D_{1Tq}^{\perp\Lambda}(z_\Lambda,Q)}{\sum_{q}e_{q}^{2}f_{1q}(x_B,Q)D_{1q}^{\Lambda}(z_\Lambda,Q)}\frac{\left\langle k_{\perp}^2 \right\rangle +z_\Lambda^2\left\langle p_{\perp}^2 \right\rangle  }{\left( \left\langle M^2_D \right\rangle+z_\Lambda^2\left\langle p_{\perp}^2 \right\rangle \right) ^2}\\
        &\ \ \ \ \ \ \times\frac{\left\langle M^2_D \right\rangle\boldsymbol p_{\Lambda\perp}}{z_\Lambda M_\Lambda}e^{\left\lbrace \frac{\boldsymbol p_{\Lambda\perp}^2}{\left\langle k_{\perp}^2 \right\rangle +z_\Lambda^2 \left\langle p_{\perp}^2 \right\rangle }- \frac{\boldsymbol p_{\Lambda\perp}^2}{\left\langle M^2_D \right\rangle+z_\Lambda^2 \left\langle p_{\perp}^2 \right\rangle } \right\rbrace }.
	\end{aligned}
\end{equation}
With this expression, we can estimate the magnitude of $P_\Lambda$ in SIDIS.
Taking $Q^2=5\,\rm{GeV}^2$, we plot the $P_\Lambda$ as a function of $\boldsymbol p_{\Lambda\perp}$ in Fig.~\ref{fig:polarization}.
The results are obtained for different values covered by the kinematic range of the future EicC.
To get $P_\Lambda$ dependence on the Feynman variable $x_F$, we parameterize $x_F$ as a function of Lorentz invariant variables $(x_B,z_\Lambda,Q)$ through a kinematic transformation:
\begin{eqnarray}\label{eq:xf}
\begin{aligned}
    x_{F}&=\frac{-z_\Lambda Q^{2}}{M[x_BM^{2}+(1-x_B)Q^{2}]}\left[\sqrt{Q^{2}+\frac{Q^{4}}{4x_B^{2}M^{2}}}  \right.\\
         &\left. \ \ \ +(M+\frac{Q^{2}}{2x_BM})\sqrt{\frac{4x_B^{2}M^{2}(M_{\Lambda}^{2}+\boldsymbol{p}_{\Lambda\perp}^{2})}{z_\Lambda ^{2}Q^{4}}-1}\right].
\end{aligned}
\end{eqnarray}
Using Eq.(\ref{eq:PA1},\ref{eq:xf}) and integrating over $\boldsymbol{p}_{\Lambda\perp}$, we plot the result of the $x_F$ and $z_{\Lambda}$ dependent $P_\Lambda$ in Fig.~\ref{fig:polarization}.

The statistical projection of $\Lambda/\overline{\Lambda}$ polarization is based on an integrated luminosity of 5 fb$^{-1}$, which is of same size of data sample as shown in Fig.~\ref{fig:mass}.  The statistical uncertainties follows the equation format as $\delta P \approx \frac{1}{\alpha_{\Lambda}\sqrt{N/3}}$ based on the polarization extraction procedure. 
The $\Lambda$ and $\overline{\Lambda}$ projected precision versus $p_{\Lambda\perp}$, $x_{F}$, and $z_{\Lambda}$ are also shown in Fig.~\ref{fig:polarization} together with theoretical predictions. The size of the error bars are smaller than the marker sizes, thus invisible. Depending on the statistics in different bins, the range of the errors is from 0.002 to 0.007.

\section{Summary and outlook}\label{sec:summary}
EicC is the proposed next generation nuclear physics facility which is expected to provide unique opportunities for precisely studying the 3-dimensional nucleon structure, the nuclear partonic structure, the exotic hadron states, {\it etc}. Lambda hyperon serving as a natural final state polarimetry is a powerful tool for studying nucleon spin structure and spin effect in the fragmentation process. Lambda measurements at EicC is of special importance and interest. 

Based on a conceptual design of EicC tracking system and GEANT4 simulation, we performed a detailed study for Lambda production and reconstruction. Also, taking spontaneous transverse polarization as an example, theoretical predictions are given as functions of different kinematic variables, together with statistical projections with one month's data taking at EicC. We find that measurements with EicC data taken in only one month's running based on current accelerator design could provide distinguishable constraints for different parameterizations of the fragmentation functions. 

EicC is designed to have both beams polarized, and the Lambda polarization transferred either from the lepton or the proton beam could provide important constraints on the spin dependent PDFs and FFs, in both co-linear and transverse momentum dependent framework. In the future work, more observables will be studied. What could be improved in the future also includes decay contributions from heavier particles, more realistic PID, {\it etc}. 

\section*{Acknowledgement}
We thank Tianbo Liu for the valuable discussion on the theoretical calculations. We thank the EicC tracking and heavy flavor working groups for the technical supports on detector simulation and useful suggestions on the analyses. 

\end{document}